\documentclass{aa}
\usepackage{graphicx}
\begin{document}
   \title{Interstellar extinction in the open clusters towards galactic
   longitude around $130^o$}

   \author{A.~K. Pandey\inst{1}, K. Upadhyay\inst{1}, Y. Nakada\inst{2},
   K. Ogura\inst{3},
   }

   \offprints{A.K.~Pandey~(pandey@upso.ernet.in)}

    \institute {$^1$ State Observatory, Manora Peak, Naini Tal, 263 129,
              Uttaranchal, India
              \newline $^2$ Kiso Observatory, School
              of Science, University of Tokyo, Mitake-mura, Kiso-gun, Nagano
              397-0101, Japan
              \newline $^3$ Kokugakuin University, Higashi, Shibuya-ku,
              Tokyo 150-8440, Japan}

\authorrunning{Pandey et al.}
\titlerunning{interstellar extinction in open clusters}

   \date{Received XXXX, 2002; accepted XXXX,}

   \abstract{ In this paper we present a detailed study of the intra-cluster reddening
   material in the young open clusters located around $l \sim 130^o$  using
   colour-excess diagrams and two-colour diagrams. The study   supports
   the universality of the extinction curves for $\lambda \geq \lambda_J$, whereas
   for shorter wavelengths the curve depends upon the value of the $R_{cluster}$
   (total-to-selective   absorption in the cluster region). The value of  $R_{cluster}$
   in the case of NGC 654, NGC 869 and NGC 884 is found to be normal, whereas the
   value of  $R_{cluster}$  in the  cluster regions NGC 1502 and IC 1805 indicates an
   anomalous reddening law in these regions. In the case of NGC 1502 the value of
   $R_{cluster}$ is found to be lower $(2.57\pm0.27)$ whereas  in the case of
   IC 1805 it is higher $(3.56\pm0.29)$ than the normal value of 3.1. Although the
   intra-cluster material indicates a higher value of   $R_{cluster}$  in the NGC 663
   region, the error in the estimation of  $R_{cluster}$  is too large to conclude
   anything. It is also found that the extinction process in the $U$ band in the case
   of NGC 663 seems to be less efficient, whereas in the case of NGC 869 the
   process is more efficient.

   }

   \maketitle
%
%________________________________________________________________

\section{Introduction}

   	Photometric studies are one of the most valuable and efficient tools
for determining the physical properties of open clusters (e.g. distance, age,
etc) and the interstellar matter (ISM) within the cluster as well as along the
line of sight of the cluster.  The accurate determination of the
distance of star clusters is crucial for a wide range of astronomical studies.
The interstellar extinction and the ratio of the total-to-selective extinction
$R = A_{V}/E(B-V)$ towards the cluster are important quantities that must
be accurately known to determine the distances photometrically.

  Observations of star clusters in the near-infrared bands, especially
of young clusters which are still embedded in remnants of their parental clouds,
are very useful to study the extinction behaviour in these clusters.
Observations of interstellar extinction due to the intra-cluster matter yield
fundamental information on the optical properties of the particles responsible
for extinction. The analysis of extinction curves in 20 OB stellar associations
indicates that big complexes are obscured by the same type of interstellar matter
(Kiszkurno et al. 1984). On the other hand Kre\l owski and Str\"{o}bel (1983)
compared the extinction law in two rich stellar aggregates, namely Per~OB1 and
Sco~OB2, situated in different parts of the galactic disk. They found that the
average extinction curves of the two aggregates seem to differ substantially.
The differences are large in the far-UV, which leads to the conclusion that an
average extinction law cannot be applied to all associations.  In an
analysis of two different complexes of young stars (Sco-Ori region and Perseus
region) Kre\l owski and Str\"{o}bel (1987) found different shapes of extinction
curves, supporting the assumption that the obscuring material in these two
young star complexes have different physical properties. Consequently, in the
case of young open clusters, the average
extinction curve should not be used to correct spectral or photometric
data for interstellar extinction.

      There is much evidence in the literature that indicates significant
variations in the properties of the interstellar extinction and these refer
mainly to high values of the total-to-selective extinction ratio $R$ as compared
to the normal value, 3.1, for the galactic diffuse medium (see
e.g. Pandey et al. 2000 and references therein). Whittet (1977) reported that
the value of $R$ in the galactic plane can be represented by a sinusoidal function
of the form $R = 3.08 + 0.17$ Sin$(l+175^{o})$, which indicates a
minimum value of $R$ at $l \approx 95^{o}$. However, Tapia et al. (1991) found
a value of $R=2.42\pm0.09$ for the cluster NGC 1502 ($l=143.7^{o}$), which is
significantly lower than the average galactic value of 3.1.

%-----------------------------------------
   \begin{table*}
      \caption{The details of the clusters used in the study}

      \begin{flushleft}
%      \begin{tabular}{lrr}
\begin{tabular}{lllllll}
      \hline \noalign{\smallskip}
Cluster & $l$   &  $b$  &Distance&Log age&E(B-V)& Available\\
        &       &       &(pc)    & (Yr)  &     & photometric data \\
      \noalign{\smallskip}
      \hline \noalign{\smallskip}
IC 1590 &123.13 &-06.24 & 2940 &6.54 & 0.32 & $UBVI_cJHK$     \\
Be 62   &123.99 &~01.10 & 2513 &7.22 & 0.85 & $UBV$        \\
NGC 436 &126.07 &-03.91 & 2942 &7.78 & 0.48 & $UBVI_c$       \\
NGC 457 &126.56 &-04.35 & 2796 &7.15 & 0.48 & $UBVI_c$     \\
NGC 581 &128.02 &-01.76 & 2241 &7.13 & 0.44 & $UBVI_c$      \\
Tr 1    &128.22 &-01.14 & 2520 &7.43 & 0.63 & $UBV$        \\
NGC 637 &128.55 &~01.70 & 2372 &6.96 & 0.70 & $UBV$          \\
NGC 654 &129.09 &-00.35 & 2422 &7.08 & 0.85 & $UBVR_cI_c$    \\
NGC 663 &129.46 &-00.94 & 2284 &7.13 & 0.82 & $UBVI_cJHK$    \\
Be 7    &130.13 &~00.37 & 2570 &6.60 & 0.80 & $UBV$        \\
NGC 869 &134.63 &-03.72 & 2115 &7.10 & 0.58 & $UBVR_cI_cJHK$  \\
IC 1805 &134.74 &~00.92 & 2195 &6.67 & 0.80 & $UBVR_cI_cJHK$ \\
NGC 884 &135.08 &-03.60 & 2487 &7.15 & 0.58 & $UBVR_cI_cJHK$  \\
NGC 1502&143.65 &~07.62 & 900  &7.03 & 0.74 & $UBVI_cJHK$     \\

      \noalign{\smallskip}
      \hline
      \end{tabular}
      \end{flushleft}
   \end{table*}
%----------------------------------------------------

  The above discussions indicate that the interstellar extinction shows a large range of
variability from one line of sight to another. A precise knowledge of the
spatial variability of interstellar extinction is important for the following
reasons;

(i) Since the extinction depends on the optical properties of the dust grains,
it can reveal information about the composition and size distribution of the
grains.  Consequently, variation of the extinction from one direction to other
may reveal the degree and nature of dust grain processing in the ISM
(Fitzpatrick 1999).

(ii) Since astronomical objects are viewed through interstellar dust, the
wavelength dependence of extinction is required to remove the effects of dust
obscuration from observed energy distributions.

      Uncertainties of extinction estimates limit the accuracy of dereddened
energy distributions.  Such uncertainties might be acceptably small for very
lightly reddened objects but can become important for modestly reddened objects.
The intra-cluster extinction due to the remains of the star-forming molecular
cloud decreases systematically with the age of the cluster (cf. Pandey
et al. 1990).

   While analysing the $UBVI_{c}$ data of NGC 663 we found that the $V/(U-B)$
colour-magnitude diagram (CMD) for NGC 663
cannot be explained by a normal value of the extinction ratio $E(U-B)/E(B-V)=
0.72$ (see Sect. 3). Some clusters in the direction of NGC 663 also appear
to show abnormal extinction properties. This motivated us to study the
interstellar extinction in detail towards the direction of the cluster NGC 663.

\vskip 2cm

\section{Data}
The online available database at {\bf http://obswww.unige.ch/webda/} by
Mermilliod (1995) provides an excellent compilation of the data on open clusters.
We selected a range of $120^{0}<l<150^{0}$ to study the intra-cluster
extinction in  open clusters. The online database lists 103 clusters in this
longitude range. The parameters, such as distance, $E(B-V)$, age
etc, are available for only 54 clusters. Seventeen clusters in this direction are
 quite young (log age $<$ 7.5) and are suitable to study the behaviour of
intra-cluster ISM.  $UBV$ CCD photometry for nine young clusters of the above
sample is given by Phelps and Janes (1994). A casual look at the $(U-B)/(B-V)$
two-colour diagrams (TCDs) given by Phelps and Janes (1994, hereafter PJ94)
indicates that the photometric data of 3 clusters, namely Be 62, NGC 637 and
NGC 663, do not satisfy the main-sequence (MS) curve, where apparent $UV$
excess can be seen for stars having spectral type later than $A$. PJ1994 concluded that this
may be due to $UV$ excess of pre-main-sequence (PMS) stars and this part of the
$(U-B)/(B-V)$ TCD is not suitable for reddening estimation. We will discuss
$(U-B)/(B-V)$ TCDs in ensuing sections. The present analysis has been carried out
using photometric data of 14 clusters and details are given Table 1.

\section{Extinction}
 One of the main characterstics of the diffuse interstellar matter in the Galaxy
is its irregular structure.  This makes it difficult to map the extinction as a
function of galactic longitude and distance. Young open clusters also show
differential reddening due to  the remains of the associated parental clouds. The
differential reddening decreases with the age of the clusters (cf. Pandey et al. 1990).
However, it does not show any correlation with the location of the cluster
in the galactic disk.

       Thus the extinction in star clusters arises due to two distinct sources;
(i) The general interstellar medium (ISM) in the foreground of the cluster,
and (ii)  the localized cloud associated with the cluster.
While for the former component a value of $R = 3.1$ is well accepted
(Wegner 1993, Lida et al. 1995, Winkler 1997), for
the intra-cluster regions the value of $R$ varies from 2.42
(Tapia et al.  1991) to 4.9 (Pandey et al.  2000 and references therein) or even
higher depending upon the conditions  occurring in the region.

\subsection{Extinction Curve}
Extinction has often been analyzed using a two-colour normalizations of the
form $E(\lambda-V)/E(B-V)$. In the present work the following methods are
used to derive the ratio of $E(\lambda-V)/E(B-V)$.

\noindent a) Colour excess diagrams (CEDs); method `A'

 The colour excesses of the stars in the cluster region can be obtained by
comparing the observed colours of the stars with their intrinsic colours
derived from the MKK bi-dimensional spectral classification.  For this purpose
we used the data available in the online catalogue by Mermilliod (1995).  When
multiple data points are available for a star, we used those selected by Mermilliod and given
as MKS in the online catalogue.  The colour excess  in a colour index
($\lambda - V)$ is obtained from the relation
$E(\lambda - V) = (\lambda - V) - (\lambda - V)_0$, where $(\lambda - V)_0$ is
the intrinsic colour index and $\lambda$ represents the magnitude in the
$UBVRIJHK$ pass bands.  Intrinsic colours are obtained from the MKK
spectral type-luminosity class relation given by Schmidt-Kaler (1982) for $UBV$,
by Johnson (1966) for $VRI$ converted to the Cousin's system using the relation
given by Bessel (1979), and by Koornneef (1983) for $VJHK$.

%*****************************(FIG 1)*********************
\begin{figure}
%\vspace*{-1.0in}
\includegraphics[height=4.0in,width=3.5in]{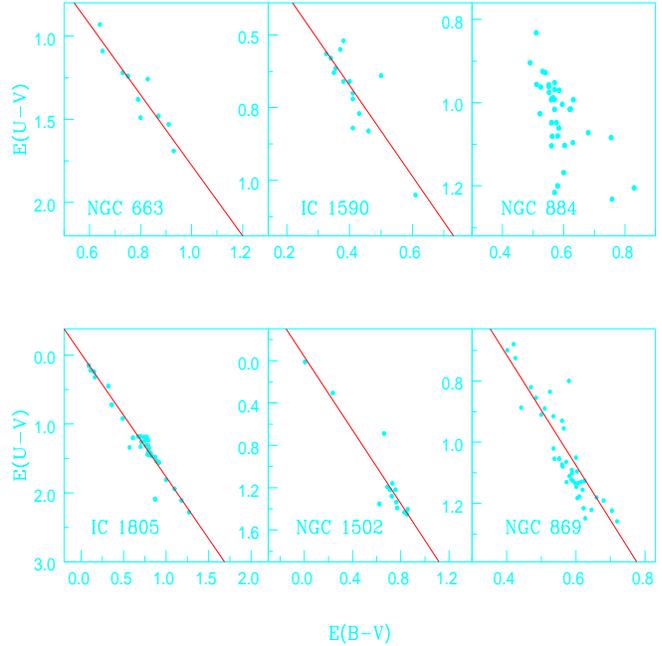}
\caption{ $E(U-V)$ vs $E(B-V)$ colour-excess diagrams. Straight line
shows a least-square fit to the data.  }
\end{figure}
%**********************************************************

The colour excesses $E(U-V), E(I-V), E(J-V), E(H-V)$ and $E(K-V)$ are shown as
a function of $E(B-V) $ in Figs.  1, 2, 3 and 4.  The stars having $H_{\alpha}$
emission features and stars  apparently lying away from the general distribution
are not included in the analysis.  A least-squares fit to the data
is shown by a straight line which gives the ratio of $E(\lambda-V)/E(B-V) $ for the stars
in the cluster region.   The slope of the
line representing the ratio $E(\lambda-V)/E(B-V) $ along with the error
is given in Table 2. In general, the least-square errors are quite large.
The reason for the large errors is mainly the small sample and the small
range in the $E(B-V)$.  For comparison the colour excess ratios for the normal
reddening law (cf. Johnson 1968, Dean et al. 1978) are also given in Table 2.

%*****************************(FIG 2)*********************
\begin{figure}
\vspace*{-2.2in}
\includegraphics[height=4.0in,width=4.5in]{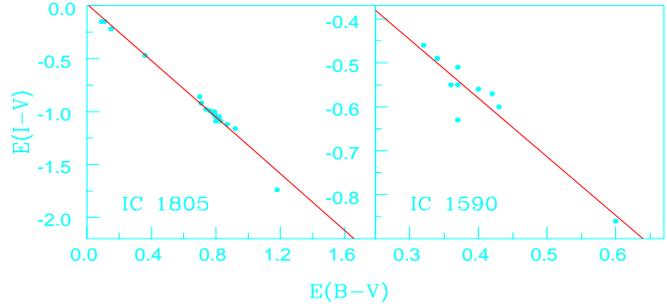}
\caption{ $E(I-V)$ vs $E(B-V)$ colour-excess diagrams. }
\end{figure}
%**********************************************************

%-----------------------------------------
   \begin{table*}
     \caption{The colour excess ratios $E(\lambda-V)/E(B-V)$ obtained from the CEDs.}
      \begin{flushleft}
\begin{tabular}{llllll}
\hline \noalign{\smallskip}
Cluster & $\frac{E(U-V)}{E(B-V)}$ & $\frac{E(I-V)}{E(B-V)}$ & $\frac{E(J-V)}{E(B-V)}$ & $\frac{E(H-V)}{E(B-V)}$& $\frac{E(K-V)}{E(B-V)}$  \\
%        &       &      &     &        &    \\
        \noalign{\smallskip}
      \hline \noalign{\smallskip}

    NGC 654 &               &               & $-2.20\pm0.26$ & $-2.47\pm0.26$ & $-2.71\pm0.27$   \\
NGC 663 & $2.12\pm0.32$ &               & $-2.28\pm0.25$ & $-2.79\pm0.30$ & $-2.99\pm0.30$  \\
NGC 869 & $1.90\pm0.13$ &               & $-1.20\pm0.46$ & $-1.75\pm0.48$ & $-1.59\pm0.50$  \\
NGC 1502& $1.74\pm0.18$ &               & $-1.87\pm0.16$ & $-2.04\pm0.23$ & $-2.17\pm0.22$  \\
IC 1805 & $1.76\pm0.08$ &$-1.34\pm0.04$ & $-2.74\pm0.30$ & $-3.30\pm0.37$ & $-3.66\pm0.38$  \\
IC 1590 & $1.56\pm0.30$ &$-1.33\pm0.17$ &                &                &              \\
Normal  & $ 1.72 $      &   $-1.25  $   & $ -2.30 $      &  $ -2.58 $     &  $ -2.78 $   \\
   \noalign{\smallskip}
      \hline
      \end{tabular}
      \end{flushleft}
   \end{table*}
%----------------------------------------------------

%**********************************(FIG 3)*********************
\begin{figure}
\includegraphics[height=4.0in,width=3.5in]{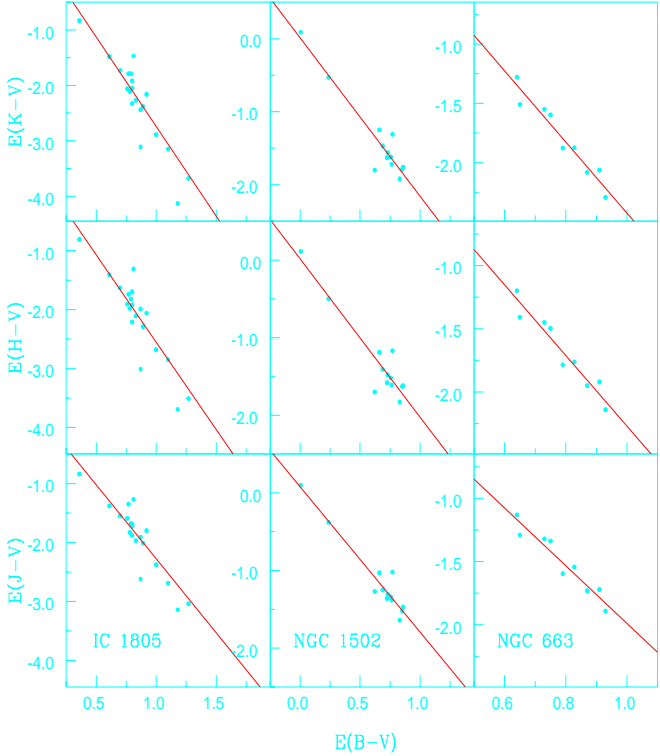}
\caption{ $E(J-V), E(H-V), E(K-V)$ vs $E(B-V)$ colour-excess diagrams
for the clusters IC 1805, NGC 1502 and NGC 663. }
\end{figure}
%***********************************(FIG 3)************************

%**********************************(FIG 4)*********************
\begin{figure}
\includegraphics[height=4.0in,width=3.5in]{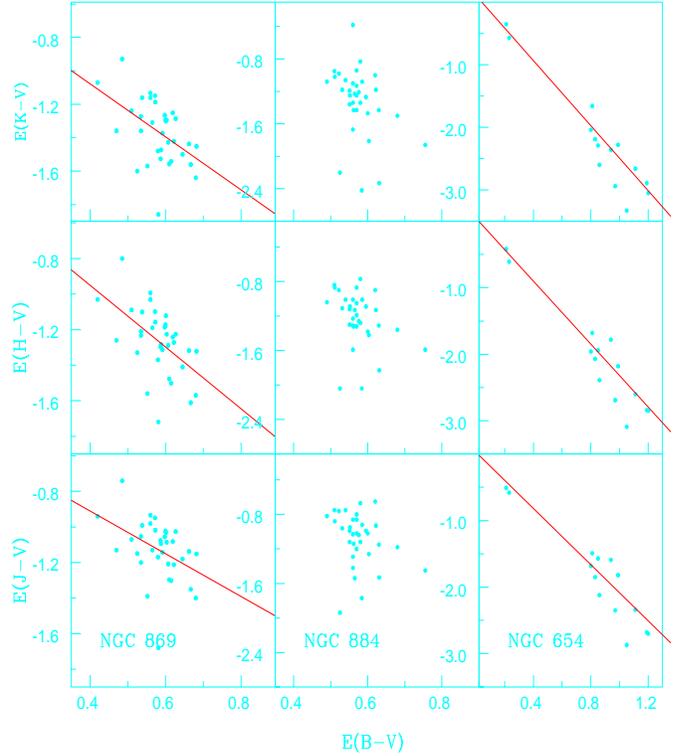}
\caption{ Same as Fig. 3 but for the clusters NGC 869, NGC 884 and NGC 654. }
\end{figure}
%***********************************(FIG 4)************************
\bigskip
\bigskip

\noindent b) Two colour diagrams (TCDs); method `B'

	In most of the cases the MKK spectral classification is available only
for a few stars of the cluster, which make the CEDs quite
noisy (see e.g.  the CEDs of NGC 654, NGC 663, NGC 869 and NGC 884).  The TCDs
of the form of $(\lambda-V)$ vs $(B-V)$, where $\lambda$ is one of the wavelength of the
broad band filters ($R,I,J,H,K,L$), provide an effective method for separating
the influence of the normal extinction produced by the diffuse interstellar
medium from that of the abnormal extinction arising within regions having a peculiar
distribution of dust sizes (cf.  Chini and Wargau 1990, Pandey et al.  2000).
On these diagrams the unreddened MS and the normal reddening vector are
practically parallel.  This makes these diagrams useless for determining the
amount of reddening, but instead, very useful for detecting anomalies in the
reddening law.  Chini and Wargau (1990) and Pandey et al. (2000) used TCDs
to study the anomalous extinction law in the clusters M16 and NGC 3603
respectively.  Figs.  5, 6, 7 and 8 show TCDs for the central region of the clusters.
We used the data of the central region of the clusters to reduce the contamination
due to field stars.
The stars  apparently lying away from the general distribution
are not included in the analysis.
The slopes of the distribution, $m_{cluster}$, are given in in Table 3.
The slopes of the theoretical MS, $m_{normal}$, on the TCDs, obtained for the
stellar models by Bertelli et al.
(1994) are also given in Table 3. The errors associated with the slopes are
significantly smaller than the errors obtained in the CEDs. The values of the
$(\lambda-V)/(B-V)$ can be converted to the ratio $E(\lambda-V)/E(B-V)$ using
the following approximate relation;

\begin{center}
  $\frac{E(\lambda-V)}{E(B-V)}=\frac{m_{cluster}}{m_{normal}}\times[\frac{E(\lambda-V)}{E(B-V)}]_{normal}$.

\end{center}

%*****************************(FIG 5)*********************
\begin{figure}
\vspace*{-0.8in}
\includegraphics[height=4.0in,width=3.5in]{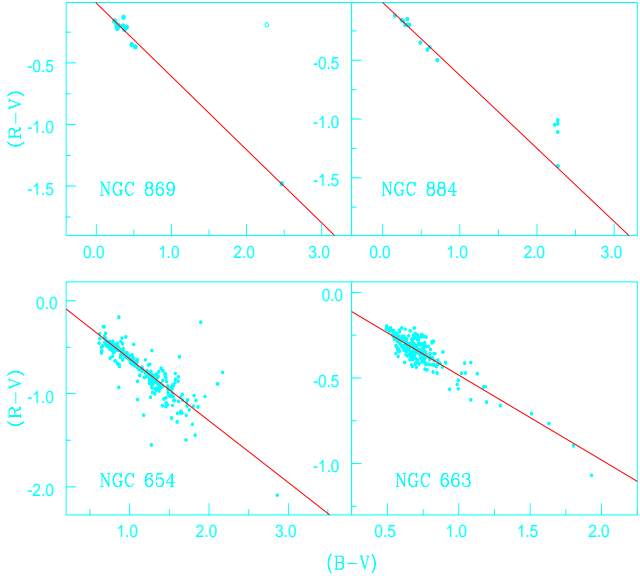}
\caption{ $(R-V)$ vs $(B-V)$ two-colour diagram. The data point shown
by open circles is not included in the least-squares fit (see text).}
\end{figure}
%*****************************(FIG 5)***********************

%**********************************(FIG 6)*********************
\begin{figure}
 \vspace*{-0.8in}
\includegraphics[height=4.0in,width=3.5in]{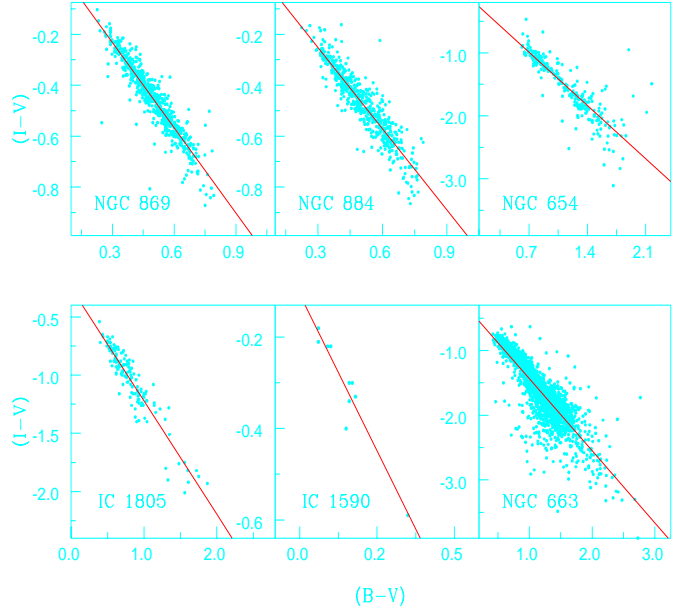}
\caption{ $(I-V)$ vs $(B-V)$ two-colour diagram.}
\end{figure}
%***********************************(FIG 6)************************

%**********************************(FIG 7)*********************
\begin{figure}
\includegraphics[height=4.0in,width=3.5in]{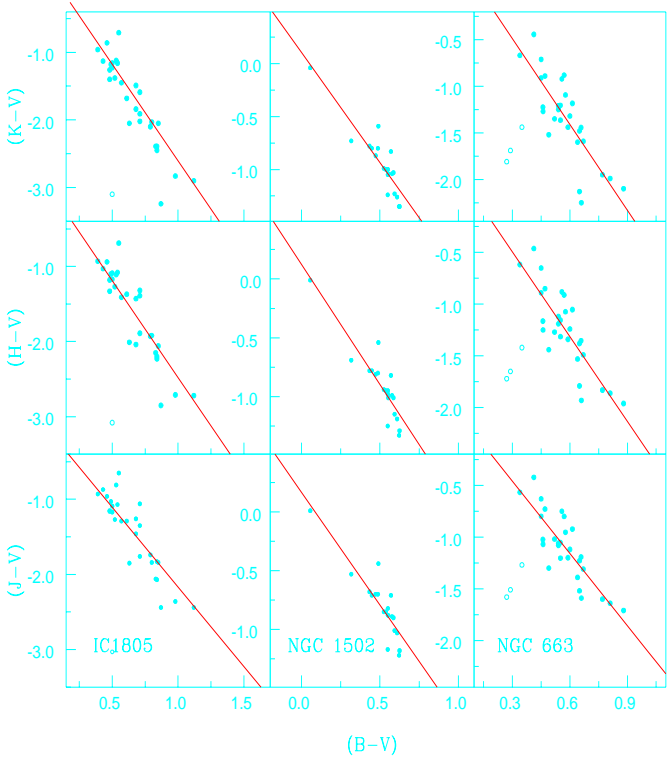}
\caption{ $(J-V), (H-V), (K-V)$ vs $(B-V)$ two-colour diagrams
for the clusters IC 1805, NGC 1502 and NGC 663. The data point shown
by open circles are not included in the least-squares fit (see text).}
\end{figure}
%***********************************(FIG 7)************************

%**********************************(FIG 8)*********************
\begin{figure}
\includegraphics[height=4.0in,width=3.5in]{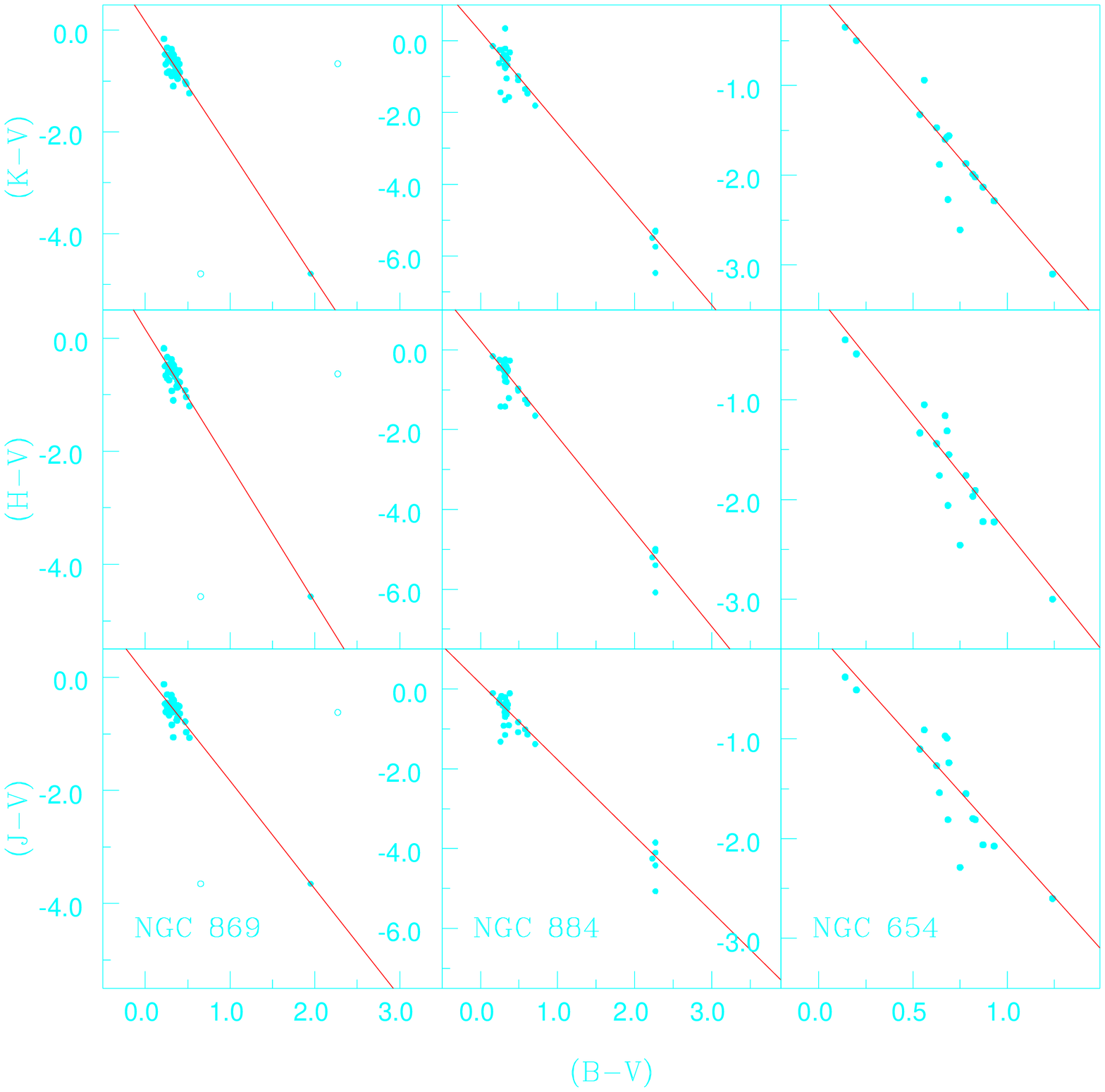}
\caption{Same as Fig. 7 but for the clusters NGC 869, NGC 884 and NGC 654.}
\end{figure}
%***********************************(FIG 8)************************
 %**********************************(FIG 9)*********************
\begin{figure}
\vspace*{-1.0in}
 \includegraphics[height=4.5in,width=5.0in]{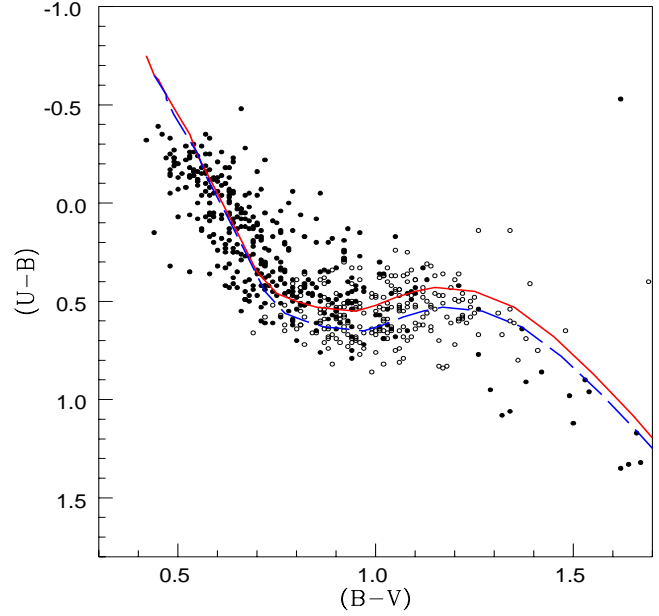}
\caption{ $(U-B)/(B-V)$ two-colour diagram for the cluster NGC 663 .
The continuous curve represents the MS shifted along $E(U-B)/E(B-V)=0.60$, whereas
dashed curve shows the MS shifted along a normal reddening vector. Star having $V\leq16.0$
are shown by filled circles.}
\end{figure}
%***********************************(FIG 9)************************
%-----------------------------------------
   \begin{table*}
     \caption{The  slopes of the distribution of stars obtained from the $(\lambda-~V)/(B-~V)$
     TCDs.}
      \begin{flushleft}
\begin{tabular}{llllll}
\hline \noalign{\smallskip}
%Cluster & $(R-V)/$     &  $(I-V)/$    & $(J-V)/$      & $(H-V)/$      & $(K-V)/$  \\
%        & $(B-V)$      &  $(B-V)$     & $(B-V)$       & $(B-V)$       & $(B-V)$   \\
Cluster & $\frac{(R-V)}{(B-V)}$ & $\frac{(I-V)}{(B-V)}$ & $\frac{(J-V)}{(B-V)} $ & $\frac{(H-V)}{(B-V)} $& $\frac{(K-V)}{(B-V)}$  \\
%        &     &      &     &      &      \\
        \noalign{\smallskip}
      \hline \noalign{\smallskip}

NGC 654 & $-0.61\pm0.02$ & $-1.20\pm0.04$ & $-2.16\pm0.25$ & $-2.39\pm0.23$ & $-2.50\pm0.25$ \\
NGC 663 & $-0.50\pm0.02$ & $-1.13\pm0.03$ & $-2.33\pm0.30$ & $-2.73\pm0.35$ & $-3.07\pm0.43$  \\
NGC 869 & $-0.60\pm0.02$ & $-1.12\pm0.02$ & $-1.90\pm0.10$ & $-2.41\pm0.10$ & $-2.53\pm0.11$  \\
NGC 884 & $-0.62\pm0.02$ & $-1.06\pm0.02$ & $-1.96\pm0.08$ & $-2.44\pm0.08$ & $-2.59\pm0.10$ \\
NGC 1502&                &                & $-1.92\pm0.24$ & $-2.04\pm0.24$ & $-2.08\pm0.24$  \\
IC 1590 &                & $-1.36\pm0.09$ &                &                &               \\
IC 1805 &                & $-0.98\pm0.04$ & $-2.12\pm0.15$ & $-2.58\pm0.17$ & $-2.73\pm0.18$  \\
Normal  & $ -0.55 $      & $ -1.1 $       &  $ -1.96 $     &  $ -2.42 $     & $ -2.60 $    \\

   \noalign{\smallskip}
      \hline
      \end{tabular}
      \end{flushleft}
   \end{table*}
%----------------------------------------------------

\medskip

\noindent c) $(U-B)/(B-V)$ two-colour diagram and colour-magnitude diagrams;
method `C'

In absence of spectroscopic observations, the $(U-B)/(B-V)$ TCD and
colour-magnitude diagrams (CMDs) are important tools to study
the interstellar reddening towards the cluster as well as intra-cluster
reddening.

 In the case of $(U-B)/(B-V)$ TCD a MS  curve (e.g.  given by Schmidt-Kaler 1982)
is shifted along a reddening vector given by the ratio of $E(U-B)/E(B-V)
\equiv X$ until a match between the MS and stellar distribution is found.
In this method the reddening vector $X$ plays an important
role.  Over the years the observational as well as theoretical studies have used
( or `misused', according to Turner 1994) a universal form for the mean galactic
reddening law. However, theoretical as well as observational estimates for the reddening
vector $X$ show a range from 0.62 to 0.80 (cf.  Turner 1994).  Recently DeGioia Eastwood
et al. (2001) also preferred a value of $X$=0.64 in the case of Tr 14. The
variability of $X$ indicates the variations in the properties of
the dust grains responsible for the extinction. High values of $X$ imply a
dominance by dust grains of small cross sections while the small values of
$X$ indicate a dominance of dust grains of larger cross section
(cf. Turner 1994). Turner  (1994) raised a question about the so called reddening-free
parameter $Q ~[=(U-B) - 0.72 (B-V)]$; how can $Q$ be reddening-free when
the reddening vector $X$ is different from 0.72?

	As we have mentioned earlier, the distribution of stars in the  $V/(U-B)$
CMD of NGC 663 cannot be explained by a normal value of $X$ (i.e. 0.72). The
$(U-B)/(B-V)$ TCD and the $V/(B-V)$, $V/(U-B)$ CMDs for the central region
of the cluster NGC 663 are shown in Figs. 9 and 10 respectively. The data have
been taken using the 105 cm Schmidt telescope of the Kiso  Observatory, Japan
(for details see Pandey et al. 2002).  In Fig. 9, where the dashed curve shows the
intrinsic MS by Schmidt-Kaler (1982) shifted along $X=0.72$,  we find a
disagreement between the observations and the MS at $(B-V)\sim 0.90$.  PJ94 explained
the disagreement due to the presence of pre-main-sequence (PMS) stars having
$UV$ excess. In Fig. 9 stars with $V\leq16$ are shown by filled circles. The
apparent distance modulus for NGC 663 is $(m-M_V)$=14.4 (cf. Pandey et al. 2002);
the stars with $V=16$ (i.e. $M_V$=1.6) will have mass $\sim 2.5 M_{\odot}$.
Since the cluster has an age of $\sim 10^7$ yr, the stars of $\sim 2.5 M_{\odot}$
should have reached the MS and no longer be PMS stars. The TCD for these stars
also does not support a normal value of  $X$.  We find that the MS shifted along a
reddening vector of 0.60 and $E(B-V)$=0.68 explains the observations satisfactorily.

	The  value of $X$ in the NGC 663 cluster region can further
be checked by comparing the theoretical zero-age-main-sequence (ZAMS) with the
observed stellar distribution in the $V/(B-V)$ and $V/(U-B)$ CMDs. Once the
reddening is known from the $(U-B)/(B-V)$ CCD, the ZAMS is shifted to match the
blue envelope of the observed stellar distribution in the $V/(B-V)$ and
$V/(U-B)$ CMDs. The ZAMS fitting for $E(B-V)$=0.68 and apparent distance modulus
$(m-M_V)$=14.4 is shown in Fig. 10a. Figure 10b shows $V/(U-B)$ CMD where ZAMS,
shifted for $E(U-B)$=0.49 (corresponding to the normal reddening vector $X=0.72$) and
$(m-M_V)$=14.4, is shown by a dashed line, which clearly shows disagreement with the
distribution of the stars. The ZAMS for $E(U-B)$=0.40 (corresponding to
$X=0.60$ and $E(B-V)=0.68$) nicely fits the blue envelope of the distribution.
This  supports an anomalous value $X$=0.60 for the slope in the NGC 663
cluster region.

%*****************************(FIG 10)*********************
\begin{figure}
\vspace*{-0.8in}
\includegraphics[height=4.0in,width=3.5in]{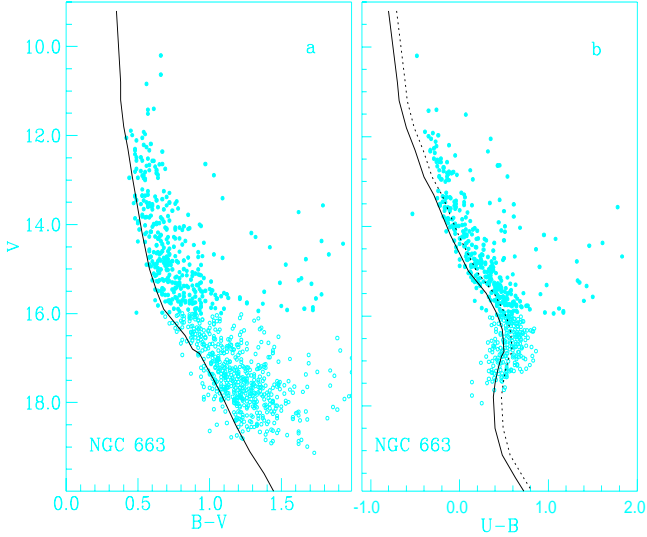}
\caption{ $V/B-V$ (the left panel), $V/U-B$ (the right panel) CMDs for the cluster
NGC 663. Star having $V\leq16.0$ are shown by filled circles.
The continuous curve in left panel represents the ZAMS fitting for
$E(B-V)$=0.68 and apparent distance modulus $(m-M_V)$=14.4.  The continuous
and dashed curves in right panel shows ZAMS shifted for $E(U-B)$=0.40  and
$E(U-B)$=0.49 respectively.  For details see text. }
\end{figure}
%******************************(FIG 10)****************************

  $(U-B)/(B-V)$ TCDs and $V/(B-V)$ CMDs were used to find out the value of
$X$ in all of the 14 clusters examined in the present study. The $(U-B)/(B-V)$
TCD for 3 clusters namely Be 62, NGC 436 and NGC 637 is shown in Fig. 11.
The values of the reddening vector $X$
obtained from the TCD/CMDs are given in Table 4. The uncertainty in the
estimated value of $X$ arises due to uncertainties in the intrinsic colours (i.e. ZAMS), uncertainties
associated with the observations and also uncertainties associated with the
visual fit of the ZAMS to the observations. The typical total  uncertainty in
the reported values of $X$ is estimated to be $\sim$ 0.05.
  %-----------------------------------------
   \begin{table}
      \caption{The value of the reddening vector $X=E(U~-~B)/E(B-V)$ obtained from the TCD/CMDs}
      \begin{flushleft}
      \begin{tabular}{ll}
       \hline \noalign{\smallskip}
       Cluster & $X$   \\
        &       \\
              \noalign{\smallskip}
      \hline \noalign{\smallskip}
IC 1590 & 0.72 \\
Be 62   & 0.60 \\
NGC 436 & 0.84  \\
NGC 457 & 0.72  \\
NGC 581 & 0.72  \\
Tr 1    & 0.72 \\
NGC 637 & 0.53 \\
NGC 654 & 0.72 \\
NGC 663 & 0.60 \\
Be 7    & 0.72  \\
NGC 869 & 0.95 \\
IC 1805 & 0.72 \\
NGC 884 & 0.72 \\
NGC 1502& 0.76 \\
      \noalign{\smallskip}
      \hline
      \end{tabular}
      \end{flushleft}
   \end{table}
%----------------------------------------------------

\section{The Value of $R$}
Table 4 indicates that the clusters Be 62, NGC 637 and NGC 663 show a smaller
value whereas the clusters NGC 869 and NGC 436 show a higher value for $X$.
The remaining 8 clusters show a normal value for $X$.
The data given in Tables 2,3 and 4 are used to estimate the weighted
(according to associated errors) mean value of the colour excess ratios
$E(\lambda-V)/ E(B-V)$ and the ratios are given in Table 5.

 We define a parameter $r$ which is the ratio of $[E(\lambda-V)/
E(B-V)]_{cluster}$ (the ratio of colour excesses in the cluster region) and
$[E(\lambda-V)/E(B-V)]_{normal}$ (the ratio of colour excesses for the normal
reddening law). The values of $E(\lambda-V)/E(B-V)$ given in Table 5 are used
to obtain the ratio $r$ for the cluster region and resultant value of $r$ is
given in Table 6. For the clusters where the data from U band to K
band are available,$r$ is plotted as a function of $\lambda^{-1}$ in Fig. 12.

%-----------------------------------------
   \begin{table*}
      \caption{Weighted mean value of the colour excess ratios $E(\lambda-V)/E(B-~V).$}
      \begin{flushleft}
\begin{tabular}{lllllll}
       \hline \noalign{\smallskip}

%Cluster & $E(U-V)/$     & $E(R-V)/$       &  $E(I-V)/$     & $E(J-V)/$      & $E(H-V)/$      & $E(K-V)/$  \\
%        & $E(B-V)$      & $E(B-V)$        &  $E(B-V)$      & $E(B-V)$       & $E(B-V)$       & $E(B-V)$   \\
Cluster & $\frac{E(U-V)}{E(B-V)}$&$\frac{E(R-V)}{E(B-V)}$ & $\frac{E(I-V)}{E(B-V)}$ & $\frac{E(J-V)}{E(B-V)}$ & $\frac{E(H-V)}{E(B-V)}$& $\frac{E(K-V)}{E(B-V)}$  \\
%        &       &      &     &        &    \\
              \noalign{\smallskip}
      \hline \noalign{\smallskip}
NGC 654 & $1.72\pm0.07$ & $-0.65\pm0.03$  & $-1.35\pm0.07$ & $-2.18\pm0.25$ & $-2.42\pm0.24$ & $-2.60\pm0.26$  \\
NGC 663 & $1.60\pm0.07$ & $-0.55\pm0.02$  & $-1.28\pm0.04$ & $-2.43\pm0.29$ & $-2.84\pm0.33$ & $-3.08\pm0.36$  \\
NGC 869 & $1.95\pm0.07$ & $-0.64\pm0.03$  & $-1.28\pm0.02$ & $-2.16\pm0.16$ & $-2.54\pm0.15$ & $-2.65\pm0.17$  \\
NGC 884 & $1.72\pm0.07$ & $-0.66\pm0.03$  & $-1.21\pm0.02$ & $-2.29\pm0.09$ & $-2.61\pm0.09$ & $-2.77\pm0.11$ \\
NGC 1502& $1.76\pm0.07$ &                 &                & $-1.97\pm0.19$ & $-2.14\pm0.24$ & $-2.20\pm0.24$  \\
IC 1805 &    $1.72\pm0.07$ &                 & $-1.25\pm0.05$ & $-2.55\pm0.22$ & $-2.86\pm0.23$ & $-3.07\pm0.24$  \\
Normal  & $ 1.72 $      & $-0.60$         & $-1.25  $   & $ -2.30 $      &  $ -2.58 $     &  $ -2.78 $   \\

      \noalign{\smallskip}
      \hline
      \end{tabular}
      \end{flushleft}
   \end{table*}
%----------------------------------------------------

	Fig. 12 indicates that the extinction in most of the cluster regions
 seems to be normal at $\lambda > \lambda_I $, except for the clusters NGC 1502
(where the colour excess ratios for $\lambda > \lambda_I$ are less than the
normal one) and NGC 1805 (where the colour excess ratio at
$\lambda \geq \lambda_J$ are higher than the normal one). Recent studies
support a universality of the extinction curves for $\lambda > \lambda_I$
(see e.g. Cardelli et al. 1989,  He et al. 1995). It is suggested that the
normalization should be done using the $E(V-K)$ instead of $E(B-V)$
(Tapia et al. 1991) because the $E(V-K)$ does not depend on properties
like chemical composition, shape, structure, degree of alignment of
interstellar dust (cf. Mathis 1990 and references therein).

 Cardelli et al. (1989) found that the mean $R$ dependent extinction law can
be represented by the following relation

\begin{center}
  $\frac{A_\lambda}{A_V}=a_\lambda + \frac{b_\lambda}{R}$   \qquad \qquad (1)  \end{center}

\noindent where $a_\lambda$ and $b_\lambda$ can be obtained from the relations given
by  Cardelli et al. (1989). The above relation can be written, in terms of
$\frac{E(\lambda-V)}{E(B-V)}$,  as

\begin{center}
  $\frac{E(\lambda-V)}{E(B-V)}=R(a_\lambda-1) + b_\lambda$     \qquad \qquad (2)

\end{center}
 %*****************************(FIG 11)*********************
\begin{figure*}
\vspace*{-2.5in}
\includegraphics[height=5.0in,width=6.5in]{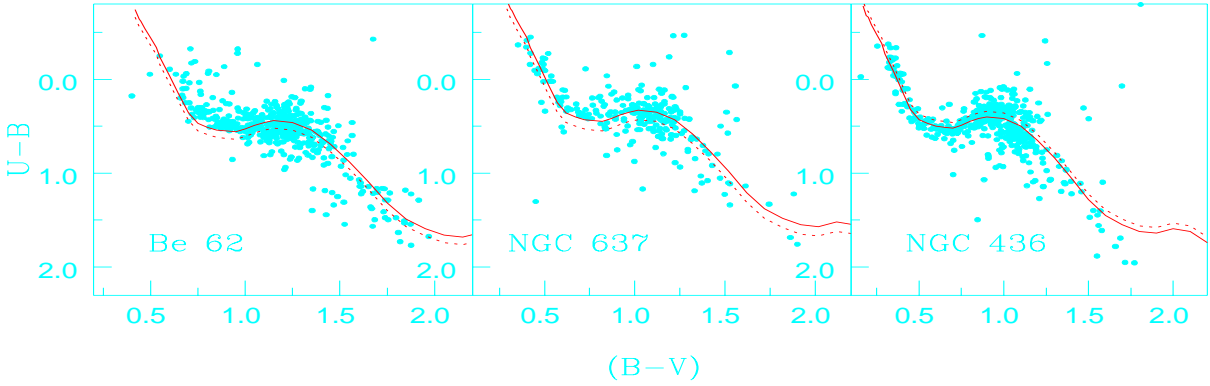}
\caption{ CCDs for the clusters Be 62, NGC 436 and NGC 637.  The continuous
curve represents the MS shifted along the reddening vector given in Table 4 and
dashed  curve shows the MS shifted along a normal  reddening vector. }
\end{figure*}
%*****************************(FIG 11)*****************************

	The ratio of total-to-selective extinction towards the cluster direction
`$R_{cluster}$' is derived using the eqn. (2).  The value of $\frac{E(\lambda-V)}{E(B-V)}$
 ${(\lambda \geq \lambda _J)}$
 given in Table 5 is used to estimate the value of $R_{cluster}$  and
the results estimate for $R_{cluster}$ are  given Table 6. The clusters that have a
broad spectrum of data are discussed below.

%-----------------------------------------
   \begin{table*}
      \caption{Mean value of the ratio $r$ as a function of wavelength.}
      \begin{flushleft}
\begin{tabular}{lllllllll}
       \hline \noalign{\smallskip}
\quad\quad $\lambda^{-1} (\mu m^{-1})$& 2.90 & 1.56 & 1.25 & 0.80 & 0.61 & 0.45                          & $R_{cluster}$  \\
Cluster     &  $U$       &  $R_C$      &$I_C$        &$J$          &$H$  &$K$       &   \\

              \noalign{\smallskip}
      \hline \noalign{\smallskip}
NGC 654     &$1.00\pm0.04$&$1.08\pm0.05$&$1.08\pm0.06$&$1.07\pm0.14$&$0.98\pm0.09$&$0.95\pm0.03$&$2.97\pm0.30 $\\
NGC 663     &$0.93\pm0.04$&$0.92\pm0.04$&$1.02\pm0.03$&$1.07\pm0.11$&$1.09\pm0.11$&$1.10\pm0.11$&$3.50\pm0.40 $ \\
NGC 869     &$1.19\pm0.02$&$1.07\pm0.05$&$1.02\pm0.02$&$0.98\pm0.05$&$1.02\pm0.05$&$1.00\pm0.04$&$3.04\pm0.20 $ \\
NGC 884     &$1.00\pm0.04$&$1.02\pm0.05$&$0.97\pm0.02$&$0.97\pm0.03$&$0.99\pm0.02$&$0.98\pm0.03$&$3.19\pm0.12$  \\
NGC 1502   &$1.02\pm0.03$&                        &                        &$0.85\pm0.05$&$0.83\pm0.05$&$0.81\pm0.04$&$2.57\pm0.27$  \\
IC 1805     &$1.00\pm0.03$&            &$1.00\pm0.02$&$1.12\pm0.09$&$1.14\pm0.08$&$1.16\pm0.07$&$3.56\pm0.29$  \\

      \noalign{\smallskip}
      \hline
      \end{tabular}
      \end{flushleft}
   \end{table*}
%----------------------------------------------------
%*****************************(FIG 12)*********************
\begin{figure}
\vspace*{-0.5in}
\includegraphics[height=5.0in,width=3.5in]{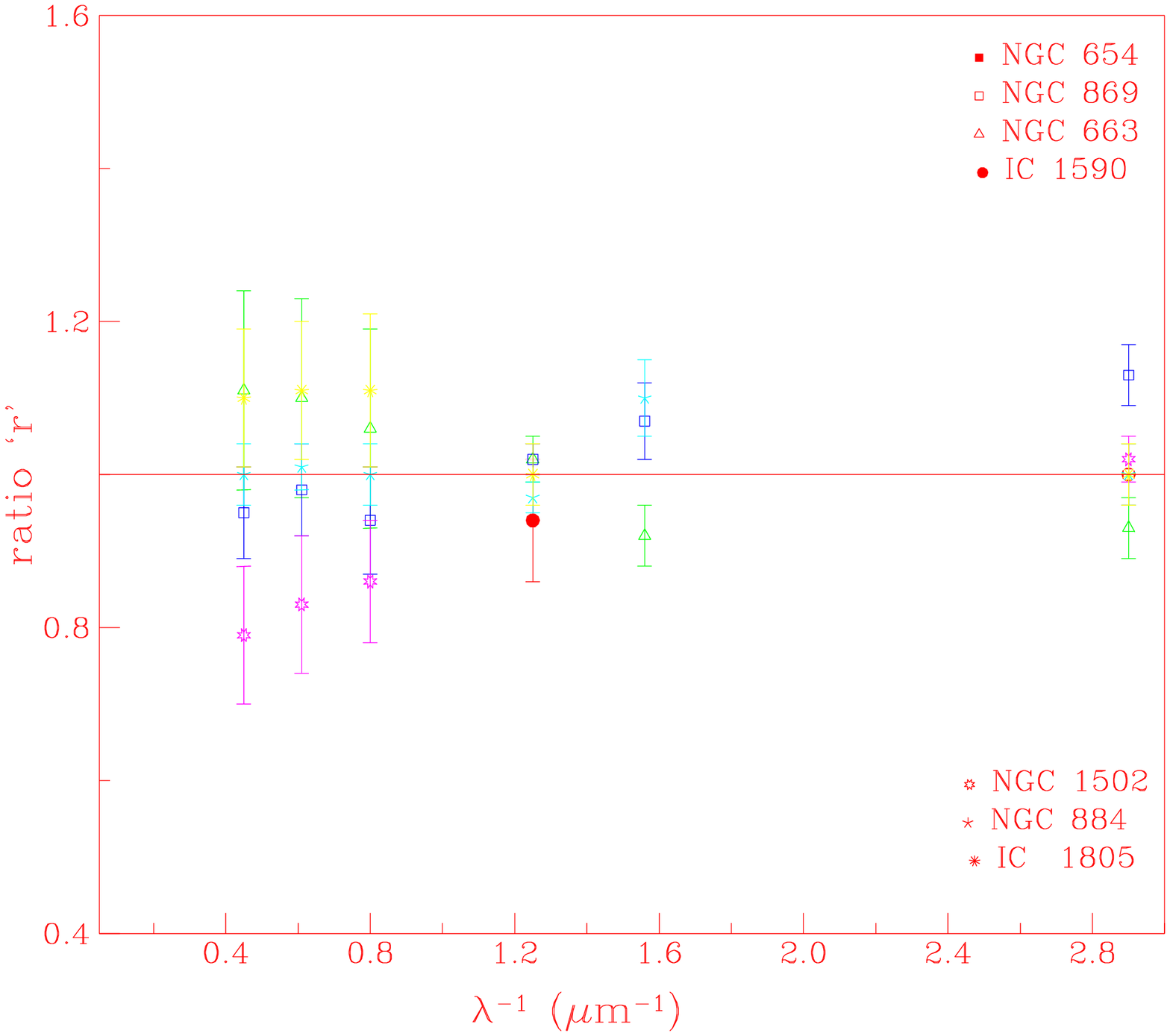}
\caption{The ratio $r$ as a function of $\lambda^{-1}$. }
\end{figure}
%*******************************(FIG 12)***************************

\bigskip

\noindent NGC 654

	Sagar and Yu (1989) concluded that at wavelengths greater than 5500
\AA, the extinction is normal. The presence of unusually well aligned
interstellar grains indicated by the polarization measurements seems to
increase the extinction in the $U$ and $B$ bands slightly (Sagar and Yu 1989).
In the present work we find a rather normal extinction law.The value of
$R_{cluster}$ is $\sim 2.97 \pm 0.30 (\sigma)$ which,
within the error, is close to the normal value of 3.1.
%\vfill \eject
 \bigskip

\noindent NGC 663

	Using the colour excesses $E(V-K)$ and $E(B-V)$ Tapia et al. (1991)
found weak evidence for an anomalous reddening law with a value of
$R_{cluster} = 2.73 \pm 0.20$, which is marginally lower than the normal value
of 3.1. However, they felt that the scatter in their data is too large to
conclude about the value of $R_{cluster}$. Yadav and Sagar (2001) reported
values for $E(\lambda-V)/E(B-V)$   (${(\lambda \geq \lambda _J)}$) which
are significantly smaller than the normal ones.

	The  (weighted) mean value of $R_{cluster} = 3.50 \pm 0.40 (\sigma)$
suggests a marginally anomalous reddening law in the NGC 663 cluster region but in the
opposite sense to that reported by Tapia et al. (1991) and Yadav and Sagar (2001).
More near-IR data is needed to determine  the $R_{cluster}$ in the
NGC 663 cluster region.  Here it is
interesting to mention that the behaviour of the extinction curve towards UV
also deviates from the normal one. Fig. 12 indicates a lower value for the
$E(U-V)/E(B-V)$ ratio, whereas Yadav and Sagar (2001) reported  a normal
value for this ratio.  They supplemented their data with the photometric
spectral types which are based on the $Q$ method, where they adopted
$E(U-B)/E(B-V)$=0.72. Presumably a dominance of photometric spectral
determination forced the ratio of $E(U-B)/E(B-V)$ to a normal value.

\bigskip

\noindent NGC 869 and 884 ($h$ and $\chi$ Persei)

	 From the extinction curve analysis Johnson (1965) found a value of
$R_{cluster} = 3.0$  in the NGC 869 and NGC 884 cluster region. Tapia et al. (1984)
also reported a normal reddening law in the cluster region. Recently
Yadav and Sagar (2001) found that the $E(\lambda-V)/E(B-V)$ ratios for $\lambda
\geq \lambda_J$ are smaller than the normal ones.

	The colour excess diagrams (method `A') of NGC 869 indicates somewhat
lower values for the ratios $E(J-~V)/E(B-V)$ and $E(K-V)/E(B-V)$ but the
errors are large, whereas in the case of NGC 884 the errors in the estimation
of colour excess ratios are too large. The reason for the large errors is a
small range in the $E(B-V) (\sim 0.3$ mag). Because of the large errors we have
not used the colour excess ratios obtained from the CEDs of NGC 869 in the further
analysis. The TCDs  (method `B') indicate that the ratio of colour excesses
$E(\lambda-V)/E(B-V)$ in both the clusters for $\lambda \geq \lambda_I$ are
perfectly normal. It is interesting to note that the CED for NGC 869 yields
$E(U-V)/E(B-V) = 1.90 \pm 0.13$. The $V/(B-V)$ and $V/(U-B)$ CMDs also seem to
support the ratio of $E(U-V)/E(B-V) = 1.9$. In a recent study Keller et al. (2001)
have adopted $E(U-B)/E(B-V) = 0.72$ to fit the ZAMS to the stellar distribution
on the $V/(U-B)$ CMD of the stars in the NGC 869 and NGC 884 cluster region.
We find that only in the case of NGC 884 the $V/(U-B)$ CMD supports a normal
value for the reddening vector $X$  in the cluster region.

\bigskip

\noindent NGC 1502

	Tapia et al. (1991) found the ratio $E(V-K)/E(B-V)
= 2.20$ corresponding to $R_{cluster} = 2.42\pm0.09$, which is significantly
lower than the reddening value. Yadav and Sagar (2001)
also found that the colour excess ratios $E(\lambda-V)/E(B-V)$ for $\lambda
\geq \lambda_J$ are significantly smaller than the normal ones. The colour excess
ratios for $\lambda \geq \lambda_J$ obtained in the present study are in good
agreement with those reported by Tapia et al. (1991) and Yadav and Sagar (2001).
The value of $R_{cluster}$ is estimated as $2.57\pm0.27$ which is in good
agreement with that obtained by Tapia et al. (1991).

\bigskip

\noindent IC 1805

	Various studies have been carried out to estimate the value of $R$
in the cluster region of IC 1805, but the
results are not conclusive. Johnson (1968), Ishida (1969) and Kwon and Lee
(1983) reported an anomalous reddening law in the cluster region with values
of $R_{cluster}$ $\sim 5.7, \sim 3.8$ and $\sim 3.44$ respectively. Kwon and Lee
also reported a regional variation in the value of $R_{cluster}$, with a maximum
value of $R_{cluster} = 3.82 \pm 0.15$ for stars located in the outer region
and the minimum of $3.06 \pm 0.05$ for stars located in the central region.
Sagar and Yu (1990) found that the interstellar extinction law in the direction
of most of the cluster members is normal. The colour excess ratios
$E(\lambda-V)/E(B-V)$ for $\lambda \geq \lambda_J$ obtained in the present work
indicate an anomalous reddening law in the cluster region of IC 1805.  The value
of $R_{cluster}$ is estimated as $3.56 \pm 0.29$, which also indicates an anomalous
reddening behaviour in the cluster region.

\section{The $A_\lambda/A_V$ curve }

	Extinction has normally been analyzed using a two-colour normalization
of the form $E(\lambda-V)/E(B-V)$. However, the true nature of the variability
of observed extinction may be hidden by the choice of normalization. The
quantity $A_\lambda/A_V$ reflects a more fundamental extinction behaviour than the
$E(\lambda-V)/E(B-V)$ (cf. Cardelli et al. 1989). The average colour excess
ratios given in Table 5 can be used to estimate the quantity $A_\lambda/A_V$ in
the following manner,

\smallskip

\begin{center}
       $A_\lambda/A_V =  [E(\lambda-V)/E(B-V) R_{cluster}]$ + 1

\end{center}

\noindent where $R_{cluster}$ is taken from Table 6. In Fig 13a the normalized
extinction in the form $A_\lambda/A_V$ is plotted against $\lambda^{-1}$ for the
clusters NGC 654, NGC 663, NGC 869 and NGC 884 alongwith the average extinction
law for $R_{cluster} = 3.1$ given by Cardelli et al. (1989). Fig 13a indicates
that the agreement between observations and the extinction law by Cardelli et
al. (1989) is good barring the $A_U/A_V$ values for NGC 869 and NGC 663. In the
case of NGC 869 the ratio $A_U/A_V$ is higher than the normal one
whereas in the case of NGC 663 it
is lower than the normal one. The value of $A_U/A_V$ = 1.46 in the case of
NGC 663 supports $R_{cluster} \sim 3.5$.  Whereas in the case of NGC 869 the
colour excess ratios indicate a perfectly normal reddening law. It seems
somewhat strange that in the case of these two clusters the ISM behaves
at $\lambda_U$ in a different way.

	The extinction law in the direction of two clusters IC 1502 and IC 1805
is found to be anomalous. The normalized extinction $A_\lambda/A_V$ for these two
clusters along with the $\rho$ Oph dark cloud (data taken from Martin and Whittet
1990) is plotted in Fig 13b. The effect of varying $R_{cluster}$ on the shape
of the extinction curves is quite apparent at the shorter wavelengths for
different environments of the star forming regions.

 As we have discussed the extinction in the direction of star clusters
 arises due to the general ISM in the foreground of the cluster and also due to the
 cloud associated with the cluster. Various studies using OB type single
 stars support a value of $R\sim 3.1$  for the general ISM (Wegner 1993, Lida et al. 1995, Winkler
 1997).  The minimum reddening, $E(B-V)_{min}$, towards the direction
 of the cluster is representative of reddening due to the foreground dust.
 The slopes of the distribution of stars having $E(B-V) \leq  E(B-V)_{min}$  on
 the TCDs can give information about the foreground reddening law. In
 the case of IC 1805 (and NGC 654), where $E(B-V)_{min}$ is 0.65 (0.77), we
 used stars having   $E(B-V) \leq$ 0.80 (0.85) to estimate the foreground
 reddening presuming that star having  $ 0.65 (0.77) \leq  E(B-V) \leq 0.80 (0.85)$
 are not much affected by the anomalous reddening law in the cluster region. The
 colour excess ratios $E(J-V)/E(B-V),  E(H-V)/E(B-V)$,  and $E(K-V)/E(B-V)$
 obtained are $-2.00\pm0.32 (-1.98\pm0.28),  -2.58\pm0.35 (-2.47\pm0.27)$ and
 $-2.80\pm0.39 (-2.82\pm0.36)$ respectively, which support a normal reddening law
 in front of the cluster  IC 1805.
 We further combined data of all the clusters having  stars with  reddening $E(B-V) \leq 0.50$.
 We feel that the limit of $E(B-V) \leq 0.50$ safely excludes the reddening due to
 intra-cluster matter as the  smallest $E(B-V)$ for the NGC 869 and NGC 884 is
 $\approx 0.50$ ( e.g. Uribe et al. 2002). A least-squares fit to 8 data points
 having $0.01\leq E(B-V) \leq 0.50 $ gives the colour excess ratio
 $E(J-V)/E(B-V)=-2.23\pm0.32,  E(H-V)/E(B-V)=-2.62\pm0.22$,  and
 $E(K-V)/E(B-V=-2.79\pm0.26$. These colour excess ratios also indicate
 a normal foreground reddening law towards the direction of the clusters
 used in the present study.

 Several studies have pointed to the apparent concentration of stars with
 high $R$-values in the vicinity of star forming regions. This effect has, for
 example, in the $\eta$ Carina nebula (Forte 1978, Th\'{e} and Groot 1983), M 16
 (Chini and Wargau 1990) and M 17 (Chini et al. 1980,  Chini and Wargau 1998)
 and it may be presumed to be characterstic of many more HII regions ( Winkler 1997).
 Winkler (1997)  compared the value of $R$ obtained for hottest stars in the
 Galaxy (spectral type O8 or earlier), which  can be considered as indicators of
 regions with recent star formation, and he confirms that in the majority of
 the cases the stars with large $R$ indeed seems to be in the vicinity of
 star forming regions.

 On the basis of the above discussions we presume that the anomalous extinction
 law in the direction of cluster IC 1805 is due
 to the intra-cluster material.
%*****************************(FIG 13)*********************
\begin{figure}
\vspace*{-1.0in}
\includegraphics[height=4.5in,width=3.5in]{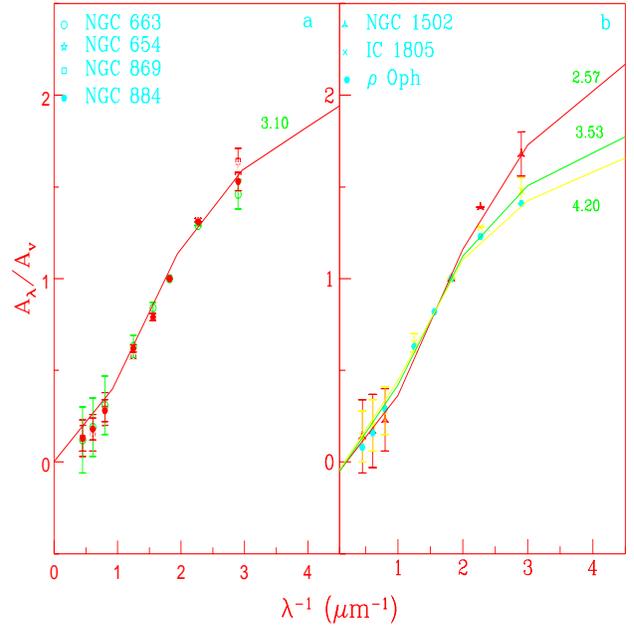}
\caption{The normalized extinction curves $A_\lambda/A_V$.  The values written
along  the curves represent the value of $R_{cluster}$}.
\end{figure}
%*******************************(FIG 13)***************************

\section{Conclusions}
In the present work we have carried out a detailed study of the
intra-cluster material inside the young open clusters within a range of
galactic longitude $123^{0} <l< 144^{0}$. We used three methods
(cf. Sect. 3.1) to derive $E(\lambda-V)/E(B-V)$ and the results agree with each
other in a few percent. It is found that  the behaviour of extinction
varies from cluster to cluster. The main results are as follows;
   \begin{enumerate}
      \item The extinction curves at shorter wavelengths depend upon
the $R_{cluster}$ while they converge for $\lambda \geq \lambda_J$.

      \item The extinction behaviour in the case of NGC 654, NGC 869 and
NGC 884 is found normal, whereas in the case of NGC 663 there is some
tendency for a higher value of $R_{cluster}$. It is interesting to mention
that in the case of NGC 663 and NGC 869 the extinction at $\lambda_U$ is found
different from the normal one. In the case of NGC 663 the extinction process in
the $U$ band seems to be less efficient, whereas in the case of NGC 869 the
process is more efficient.

      \item The cluster regions of IC 1502 and IC 1805 show anomalous reddening
laws. In the case of IC 1502 the $R_{cluster} \sim 2.57\pm0.16$ which is in
agreement with the value ($\sim2.49\pm0.09$) obtained by Tapia et al. (1991),
whereas in the case of IC 1805 the present study yields $R_{cluster}\sim3.53\pm0.25$
which is in agreement with the value ($\sim 3.44$) given by Kwon and Lee (1983).

   More near-IR data is needed for the clusters NGC 663, NGC 1502 and IC 1805,
where an anomalous value of $R_{cluster}$ is found, to study the reddening law in
detail in the these clusters.

   \end{enumerate}

\begin{acknowledgements}
      This work is partly supported by the DST (India) and the JSPS (Japan). AKP is
thankful to the staff of KISO observatory for their help during his stay there. Authors
are grateful to Prof. Ram Sagar for useful discussions.  Thanks are also due to Dr. John
Mathis for his comments which improved the contents of the paper.
\end{acknowledgements}

\end{document}